\documentstyle[12pt,epsf]{article}

\makeatletter
\@addtoreset{equation}{section}
\makeatother

\def\BbbZ{Z}
\def\half{{\textstyle{1\over2}}}

\let\a=\alpha \let\b=\beta   \let\e=\epsilon

\let\la=\label  
  
 \def\bd{\begin{document}} \def\ed{\end{document}}
\def\ds{\documentstyle} \let\fr=\frac \let\bl=\bigl \let\br=\bigr
\let\Br=\Bigr \let\Bl=\Bigl
\let\bm=\bibitem
\let\na=\nabla
\let\pa=\partial \let\ov=\overline
\newcommand{\be}{\begin{equation}}
\newcommand{\ee}{\end{equation}}
\def\ba{\begin{array}}
\def\ea{\end{array}}
\newcommand{\ho}[1]{$\, ^{#1}$}
\newcommand{\hoch}[1]{$\, ^{#1}$}
\newcommand{\bea}{\begin{eqnarray}}
\newcommand{\eea}{\end{eqnarray}}
\newcommand{\ra}{\rightarrow}
\newcommand{\lra}{\longrightarrow}
\newcommand{\Lra}{\Leftrightarrow}
\newcommand{\ap}{\alpha^\prime}
\newcommand{\bp}{\tilde \beta^\prime}
\newcommand{\tr}{{\rm tr} }
\newcommand{\Tr}{{\rm Tr} }
\newcommand{\NP}{Nucl. Phys. }
\newcommand{\tamphys}{\it
The Blackett Laboratory, Imperial College London,\\ Prince Consort Road, London SW7 2AZ}

\newcommand{\auth}{M. J. Duff\footnote{m.duff@imperial.ac.uk}}

\begin{document}
\hfill{}


\hfill{hep-th/0601134}

\vspace{24pt}

\begin{center}
{ \large {\bf String triality, black hole entropy and Cayley's
hyperdeterminant }}

\vspace{24pt}

\auth

\vspace{10pt}

{\tamphys}

\vspace{24pt}

\underline{ABSTRACT}

\end{center}

The four-dimensional $N=2$ $STU$ model of string compactification is 
invariant under an $SL(2,\BbbZ)_S \times SL(2,\BbbZ)_T \times SL(2,\BbbZ)_U$ duality 
acting on the dilaton/axion $S$, complex Kahler form $T$ and the complex structure 
fields $U$, and also under a string/string/string triality $S \leftrightarrow T \leftrightarrow U$.
The model admits an extremal black hole solution with four electric and four magnetic charges whose 
entropy must respect these symmetries. It is given by the square root 
of the hyperdeterminant introduced by Cayley in 1845. This also 
features in three-qubit quantum entanglement.

\vfill
\leftline{}

\newpage

\section{Introduction}
\la{Introduction}

An interesting subsector of string compactification to four 
dimensions is provided by the $STU$ model whose low energy limit is 
described by $N=2$ supergravity coupled to three vector multiplets. 
One may regard it as a truncation of an $N=4$ 
theory obtained by compactifying the heterotic string on 
$T^{6}$ where $S,T,U$ correspond to the dilaton/axion, complex Kahler form 
and complex structure fields respectively. It exhibits an $SL(2,\BbbZ)_S$ 
strong/weak coupling duality 
and an $SL(2,\BbbZ)_T \times SL(2,\BbbZ)_U$ target space duality. 
By string/string duality, 
this is equivalent to a Type IIA string on $K3 \times T^{2}$ with $S$ and 
$T$ exchanging roles \cite{Khurifour,Hull,Duffstrong}. Moreover, by mirror symmetry this is in turn 
equivalent to a Type IIB string on the mirror manifold with $T$ and 
$U$ exchanging roles. Hence the truncated theory has a combined 
$[SL(2,Z)]^{3}$ duality and complete $S-T-U$ triality symmetry \cite{Duff:1995sm}.
Alternatively, one may simply start with this $N=2$ theory directly as 
an interesting four-dimensional supergravity in its own right, as 
described in section \ref{STU}. 

The model admits extremal black holes solutions carrying four electric and 
magnetic charges. In section \ref{Bog} we organize these 8 
charges into the  $2 \times 2 \times 2$ {\it hypermatrix} and display 
the $S-T-U$ symmetric Bogomolnyi mass formula \cite{Duff:1995sm}. 

Associated with this hypermatrix is a {\it hyperdeterminant}, 
discussed in section \ref{Cayley}, first introduced
by Cayley in 1845 \cite{Cayley}.

The black hole entropy, first calculated in \cite{Behrndt:1996hu}, is quartic in the 
charges and must be invariant under 
$[SL(2,Z)]^{3}$ and under triality. The main result of the present paper, 
given in section \ref{Black}, is to show that this entropy is given by the square root 
of Cayley's hyperdeterminant.

The hyperdeterminant also makes it appearance in quantum information 
theory \cite{Miyake} as the measure of three qubit entanglement known as the 
{\it 3-tangle} \cite{Coffman}, which we briefly review in section \ref 
{qubit}.

\section{The STU model}
\la{STU}

Consider the three complex scalars axion/dilaton
field $S$, the complex Kahler form field $T$ and the complex structure
field $U$
\begin{eqnarray}
S&=&S_1+iS_2\nonumber\\
T&=&T_1+iT_2\nonumber\\
U&=&U_1+iU_2\ .
\end{eqnarray}
This complex parameterization allows for a natural transformation under the
various $SL(2,\BbbZ)$ symmetries.  The action of $SL(2,\BbbZ)_S$ is given by
\be
S \rightarrow \frac{aS+b}{cS+d}\ ,
\la{sl2zs}
\ee
where $a,b,c,d$ are integers satisfying $ad-bc=1$, with similar
expressions for $SL(2,\BbbZ)_T$ and $SL(2,\BbbZ)_U$.  Defining the
matrices ${\cal M}_S$, ${\cal M}_T$ and ${\cal M}_U$ via
\be
{\cal M}_S=\frac{1}{S_2}
\left(
\begin{array}{cc}
1 & S_1\\
S_1 & |S|^2
\end{array}
\right)\ ,
\label{eq:sl2mat}
\ee
the action of $SL(2,\BbbZ)_S$ now takes the form
\be
{\cal M}_S\rightarrow \omega_S{}^T{\cal M}_S\omega_S\ ,
\ee
where
\be
\omega_S=
\left(
\begin{array}{cc}
d& b\\
c & a
\end{array}
\right)\ ,
\ee
with similar expressions for ${\cal M}_T$ and ${\cal M}_U$.
We also define the $SL(2,\BbbZ)$ invariant tensors
\be
\epsilon_S=\epsilon_T=\epsilon_U=
\left(
\begin{array}{cc}
0& 1\\
-1 & 0
\end{array}
\right)\ .
\ee

Starting from the heterotic string,the bosonic action for the graviton $g_{\mu\nu}$, dilaton 
$\eta$, two-form $B_{\mu\nu}$ four $U(1)$ gauge fields $A_S^a$
and two complex scalars $T$ and $U$ is \cite{Duff:1995sm}
\begin{eqnarray}
I_{STU}&=&\frac{1}{16\pi G}\int d^4x\sqrt{-g}e^{-\eta}\Bigl[
R_g + g^{\mu\nu}\partial_{\mu}\eta\partial_{\nu}\eta
-\frac{1}{12}g^{\mu\lambda}g^{\nu\tau}g^{\rho\sigma}
H_{\mu\nu\rho}H_{\lambda\tau\sigma}\nonumber\\
&&\kern9.3em
+\frac{1}{4}\Tr(\partial{\cal M}_T{}^{-1}\partial {\cal M}_T)
+\frac{1}{4}\Tr(\partial{\cal M}_U{}^{-1}\partial {\cal M}_U)\nonumber\\
&&\kern9.3em
-\frac{1}{4}{F_S}_{\mu\nu}{}^T({\cal M}_T \times {\cal M}_U){F_S}^{\mu\nu}
\Bigr]\ .
\la{S}
\end{eqnarray}
where the metric $g_{\mu\nu}$ is related to the four-dimensional canonical 
Einstein metric 
$g^c_{\mu\nu}$ by $g_{\mu\nu}=e^{\eta}g^c{}_{\mu\nu}$ 
and where
\be
H_{\mu\nu\rho}=3(\partial_{[\mu}B_{\nu\rho]}
-\half A_{S[\mu}{}^T (\epsilon_T\times\epsilon_U){F_S}_{\nu\rho]}).
\ee
This action is
manifestly invariant under $T$-duality and $U$-duality, with
\be
{F_S}_{\mu\nu}\rightarrow
(\omega_T{}^{-1}\times\omega_U{}^{-1}){F_S}_{\mu\nu}\ , \, \qquad
{\cal M}_{T/U}\rightarrow \omega_{T/U}^T \, {\cal M}_{T/U} \, \omega_{T/U}\ ,
\ee
and with $\eta$, $g_{\mu\nu}$ and $B_{\mu\nu}$ inert.  Its equations of
motion and Bianchi identities (but not the action itself) are also
invariant under $S$-duality (\ref{sl2zs}), with $T$ and $g^c{}_{\mu\nu}$ inert and
with
\be
\left(
\begin{array}{c}
{{F_S}}_{\mu\nu}{}^a\\
{\widetilde{F}_S}{}_{\mu\nu}{}^a
\end{array}
\right)
\rightarrow     \omega_S^{-1}
\left(
\begin{array}{c}
{{F_S}}_{\mu\nu}{}^a\\
{\widetilde{F}_S}{}_{\mu\nu}{}^a
\end{array}
\right)\ ,
\ee
where
\be
{\widetilde{F}_S}{}_{\mu\nu}{}^{a}=-S_2[({\cal M}_T{}^{-1} \times {\cal
M}_U{}^{-1})(\epsilon_T \times \epsilon_U)]^a{}_b  *
{F_S}_{\mu\nu}{}^{b}-S_1 {F_S}_{\mu\nu}{}^{a}  \ ,
\ee
where the axion field $a$ is defined by
\be
\epsilon^{\mu\nu\rho\sigma}\partial_{\sigma}a=
\sqrt{-g}e^{-\eta}g^{\mu\sigma}g^{\nu\lambda}g^{\rho\tau}
H_{\sigma\lambda\tau}\ ,
\ee
and where $S=S_1+iS_2=a+ie^{-\eta}$.

Thus $T$-duality transforms Kaluza-Klein electric charges
$({F_S}^3,{F_S}^4)$ into winding electric charges $({F_S}^1,{F_S}^2)$
(and Kaluza-Klein magnetic charges into winding magnetic charges),
$U$-duality transforms the Kaluza-Klein and winding electric charge of
one circle $({F_S}^3,{F_S}^2)$ into those of the other
$({F_S}^4,{F_S}^1)$ (and similarly for the magnetic charges) but
$S$-duality transforms Kaluza-Klein electric charge $({F_S}^3,{F_S}^4)$
into winding magnetic charge $({\tilde {F_S}}^3,{\tilde {F_S}}^4)$ (and
winding electric charge into Kaluza-Klein magnetic charge). In summary
we have $SL(2,\BbbZ)_T \times SL(2,\BbbZ)_U$ and $T \leftrightarrow U$
off-shell but $SL(2,\BbbZ)_S \times SL(2,\BbbZ)_T \times SL(2,\BbbZ)_U$
and an $S$--$T$--$U$ interchange on-shell.  

One may also consider the Type IIA action  $I_{TUS}$ and the Type IIB action $I_{UST}$ obtained by 
cyclic permutation of the fields $S,T,U$. Finally, one may consider an action \cite{Behrndt:1996hu} where the $S$, $T$ and $U$ fields 
enter democratically with a prepotential
\be
F=STU
\label{eq:Haxion}
\ee
which off-shell has the full $STU$ interchange but none of the $SL(2,Z)$. All four versions are 
on-shell equivalent.

\section{The Bogomolnyi Spectrum}
\label{Bog}

Following \cite{Duff:1995sm}, it is now straightforward to write down an
$S$--$T$--$U$ symmetric
Bogomolnyi mass formula. Let us define electric and magnetic charge
vectors $\alpha_S^a$ and $\beta_S^a$ associated with the field strengths
${{F_S}}^a$ and ${\tilde {F_S}}^a$ in the standard way.
The electric and magnetic charges $Q_S^a$ and $P_S^a$ are
given by
\be {F_{S}}_{0r}^a\sim\frac{Q_S^a}{r^2} \, \qquad
*{F_{S}}_{0r}^a\sim\frac{P_S^a}{r^2}\ ,
\ee
giving rise to the charge vectors
\be
\pmatrix{\a_S^a\cr \b_S^a}=\pmatrix{  S_2^{(0)} {\cal M}_T^{-1}
   \times {\cal M}_U^{-1} & S_1^{(0)} \e_T \times \e_U  \cr 0 &
-\e_T \times \e_U }^{ab} \pmatrix{Q_S^b \cr P_S^b}.
\ee
For our purpose it is useful to define a $2 \times 2 \times 2$ array
 $a_{ijk}$ via
\be
\left(
\begin{array}{c}
a_{000}\\
a_{001}\\
a_{010}\\
a_{011}\\
a_{100}\\
a_{101}\\
a_{110}\\
a_{111}
\end{array}
\right)
=
\left(
\begin{array}{c}
-\beta_S^1\\
-\beta_S^2\\
-\beta_S^3\\
-\beta_S^4\\
\alpha_S^1\\
\alpha_S^2\\
\alpha_S^3\\
\alpha_S^4
\end{array}
\right)\ ,
\ee
transforming as
\be
a^{ijk}\rightarrow
\omega_S{}^{i}{}_{l}
\omega_T{}^{j}{}_{m}
\omega_U{}^{k}{}_{n}
a^{lmn}\ .
\ee
Then the mass formula is
\begin{equation}
m^2=\frac{1}{16}a^T({\cal M}_S{}^{-1}{\cal M}_T{}^{-1}{\cal M}_U{}^{-1}
		-{\cal M}_S{}^{-1}{\epsilon}_T{\epsilon}_U
		-{\epsilon}_S{\cal M}_T{}^{-1}{\epsilon}_U
		-{\epsilon}_S{\epsilon}_T{\cal M}_U{}^{-1})a\ .
\la{us}
\end{equation}
This is consistent with the general $N=2$ Bogomolnyi formula \cite{Ceresole}. Although all theories have the same 
mass spectrum, there is
clearly a difference of interpretation with electrically charged
elementary states in one picture being solitonic monopole or dyon
states in the other. 

This $2 \times 2 \times 2$ array $a_{ijk}$ is an example  a ``hypermatrix'', a term 
coined by Cayley in 1845 \cite{Cayley} where he also introduced a ``hyperdeterminant''.

\section{The Cayley hyperdeterminant}
\la{Cayley}

In analogy with the determinant of a $2 \times 2$ matrix $a^{ij}$

\[
{\rm det}~a_2=\frac{1}{2}\epsilon^{ij}\epsilon^{lm}a_{il}a_{jm}
\]
\be
=a_{00}a_{11}-a_{01}a_{10}
\ee
the hyperdeterminant of $a_{ijk}$ is defined to be

\[
{\rm Det}~a_3=-\frac{1}{2}\epsilon^{ii^\prime}\epsilon^{jj^\prime}\epsilon^{kk^\prime}
\epsilon^{mm^\prime}\epsilon^{nn^\prime}\epsilon^{pp^\prime}a_{ijk}a_{i^\prime j^\prime m}
a_{npk^\prime}a_{n^\prime p^\prime m^\prime}
\]
\[
=  a_{000}^2 a_{111}^2 + a_{001}^2 a_{110}^2 +
        a_{010}^2 a_{101}^2 + a_{100}^2 a_{011}^2 
\]
\[
        -2(a_{000}a_{001}a_{110}a_{111}+a_{000}a_{010}a_{101}a_{111}
\]
\[
        + a_{000}a_{100}a_{011}a_{111}+a_{001}a_{010}a_{101}a_{110}
\]
\[
        + a_{001}a_{100}a_{011}a_{110}+a_{010}a_{100}a_{011}a_{101}) 
\]
\be
        + 4 (a_{000}a_{011}a_{101}a_{110} + a_{001}a_{010}a_{100}a_{111})
        \la{cayley}
\ee
The hyperdeterminant vanishes iff the following system of equations 
in six unknowns $u^{i},v^{j},w^{k}$ has a nontrivial solution, not 
allowing any of the pairs to be both zero:

\[
a_{ijk}u^{i}v^{j}=0
\]
\[
a_{ijk}u^{i}w^{k}=0
\]
\be
a_{ijk}v^{j}w^{k}=0
\ee

Other useful identities are provided by the polynomial symmetric under 
permutation of the four indices \cite{Gibbs}:
\[
P(x_{1},x_{2},x_{3},x_{4},y_{1},y_{2},y_{3},y_{4})=
x_{1}^{2}y_{1}^{2}+x_{2}^{2}y_{2}^{2}+x_{3}^{2}y_{3}^{2}+x_{4}^{2}y_{4}^{2}
\]
\[
-4x_{1}x_{2}x_{3}x_{4}-4y_{1}y_{2}y_{3}y_{4}
\]
\be
-2x_{1}y_{1}x_{2}y_{2}
-2x_{1}y_{1}x_{3}y_{3}
-2x_{1}y_{1}x_{4}y_{4}
-2x_{2}y_{2}x_{3}y_{3}
-2x_{2}y_{2}x_{4}y_{4}
-2x_{3}y_{3}x_{4}y_{4}
\ee
which obeys
\[
P(x_{1},x_{2},x_{3},x_{4},y_{1},y_{2,}y_{3},y_{4})=(x_{1}y_{1}+x_{2}y_{2}+x_{3}y_{3}+x_{4}y_{4})^{2}
\]
\be
-4(x_{1}x_{2}+y_{3}y_{4})(x_{3}x_{4}+y_{1}y_{2})
\ee
and
\[
y_{1}^{2}P(x_{1},x_{2},x_{3},x_{4},y_{1},y_{2},y_{3},y_{4})=(x_{1}y_{1}^{2}-x_{2}y_{2}y_{1}-x_{3}y_{3}y_{1}-x_{4}y_{4}y_{1}-2x_{2}x_{3}x_{4})^{2}
\]
\be
-4(x_{2}x_{3}+y_{1}y_{4})(x_{2}x_{3}+y_{1}y_{4})(x_{2}x_{4}+y_{1}y_{3})(x_{3}x_{4}+y_{1}y_{2})
\ee
Comparison with (\ref{cayley}) yields
\be
{\rm Det}~a_{3}=P(-a_{000},a_{110},a_{101},a_{011},-a_{111},a_{001},a_{010},a_{100})
\ee
For our purposes, the important properties of the hyperdeterminant are that it is a quartic 
invariant under $[SL(2,Z)]^{3}$ and under triality.

\section{Black hole entropy}
\la{Black}
The $STU$ model admits extremal black hole solutions satisfying the 
Bogomolnyi mass formula. As usual, their entropy is given by one 
quarter the area of the event horizon. However, to calculate this 
area requires evaluating the mass not with the asymptotic values of the 
moduli, but with their frozen values on the horizon which are fixed 
in terms of the charges \cite{Ferrara}. This ensures that the entropy is 
moduli-independent, as it should be. The relevant calculation was 
carried out in \cite{Behrndt:1996hu} for the model with the $STU$ 
prepotential. The electric and magnetic charges of that paper are denoted
 $(p^0, q_0),\, (p^1, q_1),\, (p^2, q_2),\, (p^3,q_3)$ with O(2,2) 
 scalar products 
\begin{eqnarray}
p^2 &=&( p^0)^2 + ( p^1)^2 -( p^2)^2- ( p^3)^2\ ,\\
q^2 &=&( q_0)^2 + ( q_1)^2 -( q_2)^2- ( q_3)^2\ ,\\
p\cdot q &=&( p^0 q_0) + ( p^1 q_1)  +( p^2 q_2)+ ( p^3 q_3)\ .
\end{eqnarray}
In these variables, the entropy is given by
\be
S=\pi \left( W(p^\Lambda,q_\Lambda)\right)^{1/2}
\ee
where
\begin{equation} \label{ww}
 W(p^\Lambda ,q_\Lambda) =-{(p\cdot q)}^2+4\bigl (
(p^1q_1)(p^2q_2)+(p^1q_1)(p^3q_3)+(p^3q_3)(p^2q_2)\bigr )\\
 - 4 p^0 q_1 q_2 q_3 + 4q_0 p^1 p^2 p^3 \ .
\end{equation}
The function $ W(p^\Lambda ,q_\Lambda)$ is symmetric under
transformations: $
p^1\leftrightarrow p^2 \leftrightarrow p^3 $
and  $ q_1\leftrightarrow q_2 \leftrightarrow q_3. $
For the solution to be consistent we have to require $ W>0 $,
otherwise the model is not defined.

If we now make the identifications
\be
\left(
\begin{array}{c}
a_{000}\\
a_{001}\\
a_{010}\\
a_{011}\\
a_{100}\\
a_{101}\\
a_{110}\\
a_{111}
\end{array}
\right)
=
\left(
\begin{array}{c}
-p^0\\
-p^1\\
-p^2\\
-q_3\\
p^3\\
q_{2}\\
q_{1}\\
-q_{0}
\end{array}
\right)\ ,
\ee
we recognize from (\ref{cayley}) that
\be
W=-{\rm Det}~a_{3}
\ee
and hence the black hole entropy is given by
\be
S=\pi \sqrt{ -{\rm Det}~a_{3}}
\ee

Some examples of supersymmetric black hole solutions \cite{Rahmfeld1} are provided by the 
electric Kaluza-Klein black hole with $\alpha=(1,0,0,0)$ and $\beta=(0,0,0,0)$; the electric 
winding black hole with $\alpha=(0,0,0,-1)$ and $\beta=(0,0,0,0)$; 
the magnetic Kaluza-Klein black hole with $\alpha=(0,0,0,0)$ and 
$\beta=(0,-1,0,0)$; the magnetic winding black hole with 
$\alpha=(0,0,0,0)$ and $\beta=(0,0,-1,0)$. These are characterized by a 
scalar-Maxwell coupling parameter $a=\sqrt{3}$. By combining these 1-particle states, we may build up 2-, 3- and 4-particle bound states at threshold 
\cite{Rahmfeld1,Duff:1995sm}. For example $\a=(1,0,0,-1)$ and $\beta=(0,0,0,0)$ with $a=1$; 
$\a=(1,0,0,-1)$ and $\b=(0,-1,0,0)$ with $a= 1/\sqrt{3}$; $\alpha=(1,0,0,-1)$ and 
$\beta=(0,-1,-1,0)$ with $a=0$. The 1-, 2- and 3-particle states all yield 
vanishing contributions to ${\rm Det}~a_{3}$. A non-zero value is obtained for the 
4-particle example, however, which is just the Reissner-Nordstrom black 
hole.

\section{3-qubit quantum entanglement}
\la{qubit}

Interestingly enough, Cayley's hyperdeterminant also makes its 
appearance in quantum information theory.

Let the system ABC be in a pure state $|\Psi\rangle$, 
and let the components of $|\Psi\rangle$ in the standard basis be $a_{ijk}$:
\begin{equation}
|\Psi\rangle = \sum_{ijk}a_{ijk}|ijk\rangle
\end{equation}
or
\[
|\Psi\rangle = a_{000}|000\rangle+a_{001}|001\rangle+a_{010}|010\rangle+a_{011}|011\rangle
\]
\be
+a_{100}|100\rangle+a_{101}|101\rangle+a_{110}|110\rangle+a_{111}|111\rangle
\ee
In this context the $a_{ijk}$ are complex numbers rather than 
integers and the symmetry is $[SL(2,C)]^{3}$ rather than $[SL(2,Z)]^{3}$.
The three way entanglement of the three qubits A, B and C is given by 
the {\it 3-tangle} of Coffman, Kondu and Wooters \cite{Coffman}
\[
\tau_{ABC}=2|\epsilon^{ii^\prime}\epsilon^{jj^\prime}\epsilon^{kk^\prime}
\epsilon^{mm^\prime}\epsilon^{nn^\prime}\epsilon^{pp^\prime}a_{ijk}a_{i^\prime j^\prime m}
a_{npk^\prime}a_{n^\prime p^\prime m^\prime}|
\]
\be
=4 |{\rm Det}~a_{3}|
\ee
The 3-tangle is maximal for the GHZ state $|000\rangle+|111\rangle$ 
\cite{Greenberger} and vanishes for the states $p|100\rangle+q|010\rangle+r|001\rangle$.
The relation between three qubit quantum entanglement and the Cayley 
hyperdeterminant was pointed out by Miyake and Wadati \cite{Miyake}.

Thus Cayley's hyperdeterminant provides an interesting connection, at least at the level of 
mathematics, between string theory and quantum 
entanglement. Other mathematical similarities are provided by 
the division algebras \cite{Bernewig} and by twistors \cite{Levay}. 
What about physics? The near horizon geometry of the black holes is $AdS^2 \times S^{2}$ 
and one might expect a relation between the black hole entropy and the entanglement 
entropy of the conformal quantum mechanics that lives on the boundary \cite{Hawking}, 
although the nature of this particular AdS/CFT duality is not 
well-understood \cite{Michelson}. In any event, the 3-tangle is 
not the same as the entropy of entanglement \cite{Sudbery}. So the appearance of the Cayley 
hyperdeterminant in these two different contexts of stringy black hole 
entropy and 3-qubit quantum entanglement remains, for the moment, a purely 
mathematical coincidence.

\bigskip
\noindent
{\Large {\bf Acknowledgements}}

It is a pleasure to thank Chris Hull, Jerome Gauntlett, Kelly Stelle, Arkady Tseytlin and Dan Waldram
for useful conversations and Jim Liu, Anthony Sudbery, Martin Plenio and Joachim Rahmfeld for 
correspondence. I am especially grateful to Peter Levay who first drew my attention 
to Cayley's hyperdeterminant.

\bigskip

\newpage

\bibliographystyle{preprint}



\end{document}